# Best Operator Policy in a Heterogeneous Wireless Network

Soha Farhat[1,2], Abed Ellatif Samhat[1], Samer Lahoud[2], Bernard Cousin[2]
[1]Lebanese University, Ecole Doctorale des Sciences et Technologies, Hadath, Lebanon
[2]University of Rennes I – IRISA, France

*Abstract*— In this paper, we perform a business analysis of our hybrid decision algorithm for the selection of the access in a multi-operator networks environment. We investigate the ability of the operator to express his strategy and influence the access selection for his client. In this purpose, we study two important coefficients of the previously proposed cost function, *Wu* and *Wop*, and show that the value of these coefficients is not arbitrary. Simulation results show that the value of the ratio *Wu/Wop* enables a selection decision respecting operator's strategy and it affects the achieved global profit for all cooperating operators.

## I. Introduction

To cooperate, or not to cooperate: It shouldn't be anymore a difficult decision to take, for wireless network operators. Cooperation is promising and brings a lot of benefits for network operators, in terms of capacity, data rates and coverage expansion. Besides, it helps operators to decrease investment costs and brings more revenues [1]. Moreover, in a multi-operator sharing network, where competing operators share their radio access networks, operators can avoid the waste of radio resources, and defeat QoS degradation through cooperation.

However, when an operator decides to cooperate, this decision must be done with an intangible consciousness of what and how to share, and with whom to cooperate. Network operator must be aware of different characteristics of his networks and other competing operators' networks such size, geographical coverage and deployment, operators' market share, access technology, service availability, cost ...etc.

In fact, in a cooperative environment, when an operator is unable to insure satisfaction constraints to its user, he tries to give him access to the service through another network operator, thus avoiding his rejection. But, among cooperating operators, one operator may be able to offer the best QoS specifications for the user's application; while another may be able to guarantee higher profit for the user home operator. Then, the choice of the cooperating operator for user transfer may induce different profits, and it is to consider when cooperating. Additionally, competing operators can adopt different strategies when it comes to increase profits or to satisfy the user, and the priority of each goal can change with time, shared network state or the competition situation. Moreover, when the decision process of the access selection is based on a cost function as the case in our work, the strategy of the operator must be explicitly expressed and be effective, when tuned, at the selection.

In a previous work [6], a hybrid decision algorithm for the selection of the access in a multi-operator network was proposed. A cost function was used combining the performance information given by the different wireless network and the requirements of the mobile user's application, added to the resulting profit of the user exchange. The proposed cost function considered user preferences and operator's strategy, in order to guarantee the ABC user profile and a global gain for all cooperating operators. Simulation results proved the efficiency of the proposed scheme in terms of user and operator satisfaction, load balancing and network performance enhancement.

This paper extends the previous work to investigate the effect of the weights related to operator strategy [$W_u$ $W_{op}$], used in the previously proposed cost function, and show the effect of the ratio *Wu/Wop* on the selection decision and the global profit.

The rest of the paper is organized as follows: Section II presents some existing work related to business models in a multi-operator environment. Section III presents the system model and the proposed scenario to evaluate the decision cost function for a hybrid decision algorithm. Simulation results are discussed in section IV. And section V summarizes paper conclusions.

## II. Background and Related Works

In a multi-operator multi-technology wireless network, an independent Radio Resource Management (RRM) is inefficient, and a proper interworking is needed to perform a Coordinated Radio Resource Management (CRRM) in order to prevent any resource under-utilization or an overload in any Radio Access Network (RAT).

In literature, CRRM in a multi-operator context has been faced from different perspectives [1-3]. In [3] a two layered approach is adopted to improve radio resource usage and different business models are studies to show that cooperation is always beneficent for all operators. Authors introduced the concept of the Metaoperator, who acts as a third trusty party and maintains and guarantees inter-operator agreements especially for service pricing. This Metaoperator is also, responsible of the decision making process for operator selection; he communicates with the different cooperating operators and according to the information collected the decision process is triggered and the suitable serving operator is selected. As well, authors in [1] found that a third party is needed to preserve competition between operators and prevent any false information messages between them. Moreover, authors proposed that network selection is performed by either the service provider or an inter-connection provider, even though it reduces network performance due to a less tight integration.

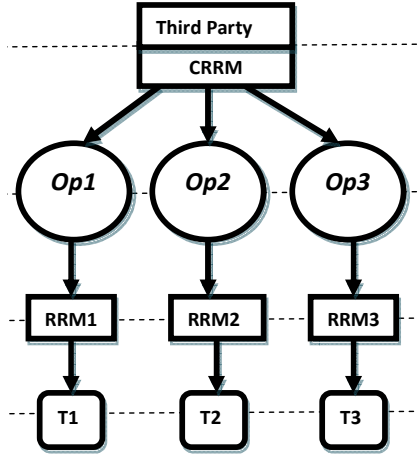

Fig. 1. Multi-operators mono-technology environment.

Consequently, in a multi-RAT multi-Operator environment, when an operator is unable to satisfy his client, he communicates with the third party to trigger the access selection process. This will trigger the RRM units of each operator to find the suitable RAT for the guest user. The third party is made to communicate with all cooperating operators and retrieve functional statistics for the selection decision in a two layered framework. Then, from different offered access networks the third trusty party will select the best RAT. Thus, finding the best RAT, in an operator centric vision, is to select the access technology guaranteeing better profits. How can we insure the selection of the best RAT? In our approach, two weights of our cost function will be tuned to control the selection.

In the following we will present the cooperating system and the decision algorithm adopted for the access selection. Then, the previously conceived cost function is described briefly and the choice of $W_u/W_{op}$ is analyzed for the selection control.

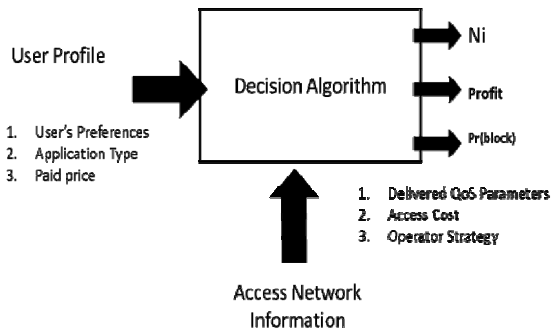

Fig. 2. Decision parameters.

### III. OPERATOR SELECTION ALGORITHM

When a group of operators decide to cooperate and share their Radio Resource, a previous establishment of inter-operator agreements is to be done and guaranteed by a third trusty party to preserve competition. In addition, a selection algorithm is to be maintained and processed in a suitable unit guaranteeing a correct decision. Fig.1 shows a system formed by three cooperating operators, each managing a single radio access network. A connecting user can be served in the network of the home operator, denoted by H-op, bound by a contract, or in the network of another operator, denoted by serving operator. When, H-op cannot admit his user or is unable to insure QoS requirements for his application, H-op intends to transfer this user to another cooperating operator to avoid his rejection. This transaction is transparent to the users.

#### A. Decision Cost Function

In our previous work [6], we proposed a cost function able to insure user's application requirements and guarantee lower service cost for the H-op. This cost function intends to minimize the distance between the score of the service network and the score of the user $|S_u-S_T|$. In addition, it permits to select the service operator who maximizes the home operator profit per exchange $(p-C_s)$. Where, $S_u$ is the score of the user combining his application QoS requirements and his paid price $p$ H-op, $S_T$ is the network score combining different QoS delivered parameters and the service price $sp$ set by the service operator for his own clients, and $C_s$ is the service cost set by the service operator for guest service. Fig.2 shows the input parameters for the decision algorithm. A group of parameters comes from the user as the application type, the paid prices and user preferences, and a group of parameters are available from each cooperating operator as the QoS delivered parameters, the service price $sp$ and the service cost $C_s$. The decision process will select the best operator by minimizing the following cost function:

$$CF = W_u * |S_u - S_T| - W_{op} * (p - C_s) \qquad (1)$$

We can point here to $W_u$, the weight determining the degree of importance for the H-op, to satisfy the user, and $W_{op}$, the weight determining the degree of importance for the H-op, to improve his profits. The choice of the ratio $W_u/W_{op}$ is not arbitrary and is not simply related to the degree of importance of satisfying the user over achieving higher profits as in AHP scale [5]. In fact, an analytical study of the proposed cost function showed that this ratio is bounded by limits, which are function of the difference between the achieved profit $(p-C_s)$ and the difference between the scores distances $|S_u-S_T|$, when candidate operators are compared pair wise. So, the value of $W_u$ and $W_{op}$ could be dynamic or static depending on the considered scenario. The limits mentioned above determines the group of values of the ratio $W_u/W_{op}$ that permits to select the operating maximizing the H-op profit, and the group of values that permits to satisfy, to the maximum, user's application requirements. To illustrate what it is described above, we consider two candidates operator for the selection, *Op1* and *Op2*, having the scores $S_{T1}$ and $S_{T2}$ and setting *Cs1* and *Cs2* for the service cost, thus having *CF1* and *CF2* as cost function, respectively:

$$\begin{cases} CF1 = W_u * |S_u - S_{T1}| - W_{op} * (p - Cs1) \\ CF2 = W_u * |S_u - S_{T2}| - W_{op} * (p - Cs2) \end{cases} \quad (2)$$

If the profits per exchange are higher when the user is served by Op1, H-op must set *CF1<CF2* to select *Op1*, thus $W_u/W_{Op}$ must fulfill the condition:

$$\frac{W_u}{W_{op}} < \frac{(p-Cs1)-(p-Cs2)}{|S_u-S_{T1}|-|S_u-S_{T2}|} = L \quad (3)$$

In other scenarios, where there are more than two candidates for the selection, the ratio $W_u/W_{op}$ will be bounded by two or more limits.

## IV. SIMULATION RESULTS

### A. System Model

We consider three cooperating operators, $Op_1$, $Op_2$ and $Op_3$, each managing a single radio access network *UMTS*, *WLAN1* and *WLAN2*, respectively. The conditions of the networks are shown in Table I. In this study, we suppose that all RATs are capable of delivering a constant value for the mean jitter $J_M$, mean delay $D_M$ and mean bit error rate $BER_M$ [4]. The normalization of the different parameters is done for each access network with respect to the service requirements. We consider one single service type for real-time application, this service requirements are determined in Table II, taking into account that bandwidth consumption of a service varies with the network technology.

TABLE I. UMTS AND WLAN NETWORK PARAMETERS [4]

| Network Technology | QoS Parameters | | | |
|---|---|---|---|---|
| | Bandwidth(Kb/s) | Jitter(ms) | Delay(ms) | BER(dB) |
| UMTS | 1700 | 6 | 19 | $10^{-3}$ |
| WLAN1 | 11000 | 10 | 30 | $10^{-5}$ |
| WLAN2 | 5500 | 10 | 45 | $10^{-5}$ |

TABLE II. APPLICATION REQUIREMENTS

| Service Type | QoS Parameters | | |
|---|---|---|---|
| | Jitter(ms) | Delay(ms) | BER(dB) |
| Real-Time | 10 | 100 | $10^{-3}$ |

### B. Effect of Wu/Wop

To study the effect of the ratio $W_u/W_{op}$ on the selection decision, we consider the system model described above with the jitter as decision parameter. Besides, without loss of generality, the service price *sp* is set to 0.9, 0.1, and 0.5 units/Kbytes for *Op1*, *Op2* and *Op3* respectively. The service cost is set equal to the service price for all cooperating operators *Cs=sp*. Consequently, a client served in his H-op network will pay for the H-op *p=sp*, and when this user is served in another S-op network he will pay *p=spi* for his H-op *Opi* and the latter will pay *Cs=spj* for the S-op *Opj*.

As described previously, when there are two candidates for selection, one operator may offer the best score and the other may offer the best profit for the H-op. Thus, an upper limit *L* bounds the values of $W_u/W_{op}$ that guarantee the selection of the operator with best profit, and all values above *L* select the best scored operator. *L* can be calculated from equation (3): so, if the H-op intends to choose the service operator with lower cost, $W_u/W_{op}$ must be chosen below *L*, else if he seeks the operator delivering the best parameters for his client, $W_u/W_{op}$ must chosen above *L*. Note that the limit *L*, for every operator, changes with the system model, the pricing scenario and the score distances (if the considered QoS parameters vary during simulation).

In this scenario, the mean jitter delivered by RATS is constant, so L is constant. And for more simplification a common limit is deduced analytically for the three cooperating operators, and without loss of generality, two values are decided for simulation $W_u/W_{op}=1/4<L$ and $W_u/W_{op}=8>L$.

*1) Effect on the selection decision:* We are interested, to present for each H-op, the service o*perator, S-op, with a minimum* score *distance (*dmin*)* to the user requirement*s*, and the *S-op generating best profit for* H-op*, in this scenario,* Table III. Simulation results for the effect of the choice of Wu/Wop on the selection are presented in Table IV. Selection results show that, a ratio below the limit L guarantee all the time that H-op selects the operator maximizing his profits. Besides, when the selection is between two operators causing losses for H-op, the selection algorithm guarantee the lower loss with

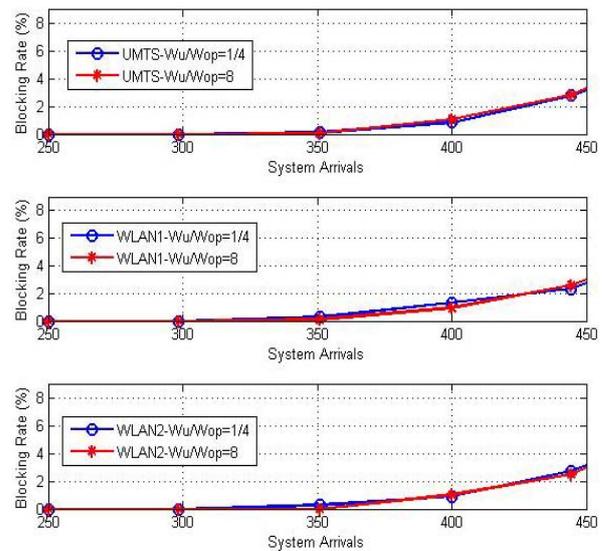

Fig. 3. Blocking probability conservation

$Wu/Wop<L$.

In fact, in this scenario where Cs=sp, we can distinguish three cases:
1. The H-op must choose between two operator inducing different profits (case of *Op1*): in this case, when *Wu/Wop* is tuned below *L*, the S-op with better profit is selected.
2. The H-op must choose between an operator with some profits and other causing him losses (case of *Op3*): in this case a value below *L* guarantees the selection without losses.
3. The H-op must choose between two operators, each having high service costs and causes losses to the H-op (case of *Op2*): in this case, when *Wu/Wop* is lower than *L*, H-op selects the S-op causing him lower loss.

And, note that for all cases, when $Wu/Wop>L$ the selection goes to the operator having the closest score to *Su*.

TABLE III. CANDIDATE OPERERATORS QUALIFICATION

| H-op | Candidate Operator | |
|---|---|---|
| | with dmin | with Best profit |
| Op1 | Op3 | Op2 |
| Op2 | Op1 | Op1 |
| Op3 | Op1 | Op2 |

TABLE IV. SERVICE OPERATOR SELECTION (%)

| Selection Direction | | | | | | | |
|---|---|---|---|---|---|---|---|
| Wu/Wop=8 (dminOp selection) | | | | Wu/Wop=1/4 (best profit Op selection) | | | |
| | Op1 | Op2 | Op3 | | Op1 | Op2 | Op3 |
| Op1 | - | 0 | 100% | Op1 | - | 100% | 0 |
| Op2 | 100% | - | 0 | Op2 | 100% | - | 0 |
| Op3 | 100% | 0 | - | Op3 | 0 | 100% | - |

Hence, results showed that the ratio *Wu/Wop* had reflected the operator strategy for the selection decision. But, it would be necessary to test if tuning the operator strategy (varying *Wu/Wop* above and below *L*) has any effect on the network performance and if it ameliorates the global profit, achieved by each operator. Results are presented briefly in the following subsectors:

*2) Effect on the network performance:* We used the blocking rate as an indicator for the network performance. Fig.3 represents this blocking probability for all operators when varying the strategy ratio *Wu/Wop*. Results showed that strategy ratio has no effect on the performance; the blocking probability is conserved within a 90 % confidence intervalle.

*3) Effect on the operator profit:* So, it is enough to have information about the service cost set by different cooperating operators and the delivered QoS score, to decide the values of *Wu* and *Wop* and insert them into the decision cost function to guarantee the selection of the best operator ( achieving best profit for H-op). But, results showed that the selection of the best operator won't always ameliorate the global profit achieved by H-op. Fig.4 represents operators' global profits variation when tuning the strategy ratio. It shows a great amelioration in profits for *Op3*, when setting $Wu/Wop<L$. In fact, *Op3* had to choose between *Op2* that brings profits and *Op1* that causes losses. So, when the strategy is changed to a value>*L* and the selection falls 100% to *Op1*, *Op3* was penalized with a lot of losses causing profit degradation.

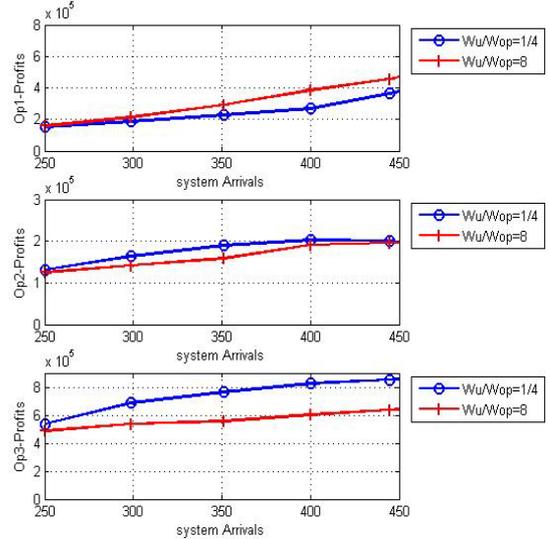

Fig. 4. Operators' global profit variation

It is the same for *Op2*, little improvement is detected since a value of $Wu/Wop<L$ permits to select the operator with lower losses. In fact, in this scenario, *Op2* offers the cheapest service, so he is always loosing when exchanging a client, in term of revenue. But, it is not the same for *Op1*. Tuning *Wu/Wop* below *L* causes lower profit, but a value>*L* ameliorates it. Indeed, the value of *Wu/Wop* affects the whole exchange process and thus, the probability that each operator acts as an S-op and the amount of added guests in each operator network, besides the direction of users' migration from each H-op and thus the charged costs. And, while the global profit is calculated from the revenues of the added clients, guest users and exchanged clients minus the cost charged for the client exchange, the value of *Wu/Wop* will affect it in some way. So, in this scenario, when *Wu/Wop* was set above *L* for all cooperating operators, *Op1* is more selected to serve *Op2* and *Op3* users (Table IV), since having better score distance, which increases hugely the amount of revenues from guest users, and ameliorates his profits. We can conclude briefly, that the value of $Wu/Wop<L$ was efficient to prevent a choice with losses and ameliorate profits, but when choosing between two profiting operators the operator must consider the resulting exchange process and the probability to achieve higher revenues.

## V. Conclusion

In this paper, we have showed that our decision algorithm for access selection gives operators the ability to exchange clients with respect to their decided strategy by tuning the ratio *Wu/Wop* above or below a calculated limit *L*. This limit *L* depends of the system model, i.e, the number of cooperating operators, the delivered QoS parameters of the different networks of the competing operators, the service price *sp* set for the clients of the different operators and finally the pricing scenario adopted for the service cost *Cs* charged to guest's H-op. Moreover, setting the ratio to a value that guarantees the selection of the most profitable operators does not always guarantee a better global profit gain. The value of the ratio *Wu/Wop* will affect the direction of users exchange and the probability that an operator acts as a service operator or. Thus, operators must be aware of the user exchange process and the best pricing scenario, to have a good prediction of the extra revenues that can be achieved and the possible profit improvement gained from cooperation. Future work will investigate the pricing scenario that can be adopted during cooperation and how to choose the best one to ameliorate profits.